\documentclass[12pt]{iopart}

\usepackage{iopams} 
\usepackage{color}
\usepackage{graphicx}
\usepackage{hyperref}

\newcommand{\TFU}[2] {\widetilde{U_\mathrm{ #1}}\left(#2\right)}
\newcommand{\TFV}[2] {\widetilde{V_\mathrm{ #1}}\left(#2\right)}

\begin{document}

\title{Storage of RF photons in minimal conditions}

\author{J.-P. Cromi\`eres \& T. Chaneli\`ere}

\address{Laboratoire Aim\'e Cotton, CNRS, Univ. Paris-Sud, ENS Cachan, Universit\'e Paris-Saclay,  91405 Orsay Cedex, France}
\ead{thierry.chaneliere@u-psud.fr}
\vspace{10pt}

\begin{abstract}

We investigate the minimal conditions to store coherently a RF pulse in a material medium. We choose a commercial quartz as memory support because it is a widely available component with a high Q-factor. Pulse storage is obtained by varying dynamically the light-matter coupling with an analog switch. This parametric driving of the quartz dynamics can be alternatively interpreted as a stopped light experiment. We obtain an efficiency of 26\%, a storage time of 209$\mu$s and a  time-to-bandwidth product of 98 by optimizing the pulse temporal shape. The coherent character of the storage is demonstrated. Our goal is to connect different types of memories in the RF and optical domain for quantum information processing. Our motivation is essentially fundamental.

\end{abstract}

%
%
%
%
%


\maketitle

\section{Introduction}

The quest of electromagnetic analog memories recently reappeared in the context of quantum processing. The different implementations have covered a broad range of the electromagnetic spectrum. On the one hand, the evolution of superconducting qubits for quantum computing motives the development of memories in the RF domain. On the other hand, the quest for quantum repeaters in long-haul secured communications requires the realization of quantum memories in the optical domain. The memory media cover different realities from high-Q optical \cite{xu2006experimental, yoshikawa2013creation} and RF resonators  \cite{PhysRevLett.112.210501_Martinis,  PhysRevLett.114.090503_Flurin} to atomic vapors or atomic-like impurities in solids \cite{heshami2016quantum}, including an hybridized version of the two \cite{grezes2015storage, PhysRevX.5.031009}. Metamaterials, whose properties can be tailored, have been also considered in the same context, both in electromagnetism \cite{nakanishi2013storage} and opto-mechanics  \cite{fiore2011storing, mcgee2013mechanical}. Whatever is the physical support, the largest Q-factors are desirable for the resonant oscillator because this translates into long storage times. Atomic vapors appear as a natural candidate for quantum storage because of the unequaled sharpness of the different transitions motivating a large community.

The problem of storing light coherently has been addressed with a fundamental point of view by Yanik {\it et al.} in \cite{yanik2005stopping} and a series of related publications. One needs to break down the time-to-bandwidth product of a resonator by modifying dynamically its bandwidth. These fruitful approach has been successfully implemented and interpreted accordingly but seems to be compartmented unnecessarily to integrated photonic structures \cite{xu2007breaking}. The argument is general and doesn't depend on the physical nature of the oscillator. To be more precise, let us consider the general partial differential equation of a driven harmonic oscillator described by its amplitude $V_ \mathrm{out}(t)$:

\begin{equation} \label{driving}
\partial^2_t U_ \mathrm{out}(t) + \omega_q^2 U_ \mathrm{out}(t) + \Gamma \partial_t U_ \mathrm{out}(t) = \Gamma \partial_t U_ \mathrm{in}(t)
\end{equation}

where $ \omega_q$ is the eigenfrequency and $\Gamma$ the damping rate. The left hand side is the familiar damped oscillator equation of motion. The right hand side represents the driving term that we choose as $\Gamma \partial_t U_ \mathrm{in}(t)$. This latter can be adjusted, depending on the coupling scheme used for the driving force (written $U_ \mathrm{in}(t)$). The oscillator response is usually described in the spectral domain by a Lorentzian response of width $\Gamma$. Following the argument of \cite{yanik2005stopping}, we will make the bandwidth $\Gamma(t)$ time-dependent in order to store the incoming driving pulse $U_ \mathrm{in}(t)$ into the oscillator. Since the argument is general, we will use a simple RLC circuit in our case. $U_\mathrm{in}(t)$ and $U_\mathrm{out}(t)$ will be the input and output voltages.

It should be noted that interesting openings have been developed to reposition the quest of optical memories in the broader framework of analog random-access devices \cite{PhysRevLett.93.233903, tucker2008optical, sorin2009optical}. Along the same lines, the connection between dynamical controlled integrated photonics  and electromagnetically induced transparency (EIT) has been made explicit \cite{tucker2005slow, xu2006experimental}, thus partially filling the gap between optics and atomic physics. Despite a common motivation, there is no unified vision for these different approaches usually investigated by adjacent communities. We wish to make a contribution in this direction by assembling different pieces of what we consider as a unique jigsaw puzzle.

We here investigate the minimal conditions to coherently store a RF pulse in a high-Q resonator. Our approach is guided by frugality. Direct storage of light (RF or optical) into matter is obtained by dynamically controlling the coupling constant between light and matter \cite{yanik2005stopping}. This fundamental argument is sometimes submerged by alternative interpretations. Stopped-light experiments based on EIT represent a continuous success story. The storage step is usually interpreted as a dynamical control of the light group velocity through the so-called dark-state polariton \cite{fleischhauer2000dark}. This analysis is perfectly valid and has the advantage to account for the different energy scales. Taking the terminology of quantum memories, the flying qubit is in the optical domain and the static qubit is the atomic spin (RF domain). The order of magnitude gap between the frequencies is filled by the Raman control field. The dark-state polariton gives a physical insight of the EIT storage beyond the complexity of the three-level structure ($\Lambda$-system) and the presence of the Raman control field. Our approach is diametrically opposite because we consider a single two-level system in the linear (perturbative) regime or in other words a single high-Q oscillator. In that case, flying and static qubit have the same frequency. Fidelity and energy are unequivocally associated. As a storage medium, we use a commercial quartz crystal whose finesse is remarkable by benefiting from years of development. The coupling to the RF field is dynamically controlled (loaded or unloaded) with an analog switch. The light pulse is shaped to optimize the storage efficiency as discussed for optical quantum memory in atomic media \cite{Gorshkov_optimal, Gorshkov_A, Gorshkov_B, 0295-5075-86-1-14007_ref18_QM, PhysRevA.83.063842_ref19_QM, Heugel2010_ref38_Cl} and also for RF pulses in a high-Q superconducting resonator  \cite{PhysRevB.84.014510_ref22_Cl, PhysRevLett.112.210501_Martinis, PhysRevLett.114.090503_Flurin}. The argument is based on a time-reversal symmetry. After optimization, we obtain an efficiency of 26\% and a storage time of 209$\mu$s (section \ref{experimental}). This latter is essentially limited by the unloaded quartz oscillator linewidth.  The coherent character of the storage is demonstrated by making interfere two independent memories (section \ref{coherent}). We will briefly conclude our proof of principle demonstration by discussing its formal analogy with EIT storage in an atomic vapor (section \ref{stopped_light}) thus bridging the gap between two apparently distinct situations.

\section{Quartz oscillator with variable coupling}

We choose a quartz as a reference oscillator for our storage demonstration. The write and read memory processes are physically implemented by a dynamical variation of the coupling. Different coupling schemes are possible: conductive, inductive and capacitive. We choose a direct conductive coupling to preserve as far as possible the characteristic spectrum of the quartz and avoid a coupling inductor or capacitor which would significantly impact the quartz response. The direct resistive coupling is simple. This latter defines a band-pass filter (see fig.\ref{fig:circuit_equivalent}) which will be the basis of our analysis.

\subsection{Band-pass filter design}

\begin{figure}[htbp]
\centering
\includegraphics[width=.5\linewidth]{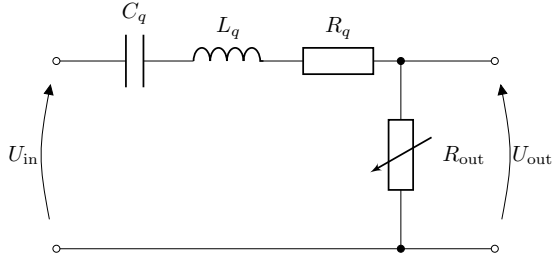}
\caption{Ideal band-pass filter with variable coupling though the load $R_\mathrm{out}$. The resonator is modeled as a series RLC with $R_q$, $C_q$ and $L_q$ respectively. $R_q$ is much smaller than the load $R_\mathrm{out}$.}
\label{fig:circuit_equivalent}
\end{figure}

We take a series RLC circuit with a direct resistive coupling (though a variable load resistor) as a toy model for a band-pass filter. The equivalent circuit is represented in fig.\ref{fig:circuit_equivalent}.

This latter is indeed described by the general equation of motion of a driven harmonic oscillator (eq.\ref{driving} with $\Gamma=R_\mathrm{out}/L_q$ and $\omega_q=1/\sqrt{L_q C_q}$, assuming $R_q\ll R_\mathrm{out}$) where $U_\mathrm{in}(t)$ and $U_\mathrm{out}(t)$ are the time-varying voltages at the input and output respectively. The transmission spectrum is described by the Lorentzian shape

\begin{equation}\label{Bode_second}
\frac{\TFU{out}{\omega}}{\TFU{in}{\omega}} = \frac{j \omega \Gamma}{j \omega \Gamma + \omega_q^2-\omega^2}
\end{equation}
where $\TFU{in}{\omega}$ and $\TFU{out}{\omega}$ are the Fourier transforms of $U_\mathrm{in}(t)$ and $U_\mathrm{out}(t)$.

This is the ideal situation that we will consider later on. To anticipate the description in section \ref{sec:storage}, we will in practice vary the load resistor from a nominal value to zero.  An incoming RF pulse $U_\mathrm{in}(t)$ transiently excites the resonator. The writing step (storage) is realized by zeroing the load resistor (uncoupled resonator). The pulse is released as $U_\mathrm{out}(t)$ by coupling back the resonator through the nominal load resistor. The storage time will be essentially limited by the resonator intrinsic quality factor (residual resistive part $R_q$ in fig.\ref{fig:circuit_equivalent}). 

\subsubsection{Experimental realization}\label{section:experimental_realization}

\begin{figure}[htbp]
\centering
\includegraphics[width=.8\linewidth]{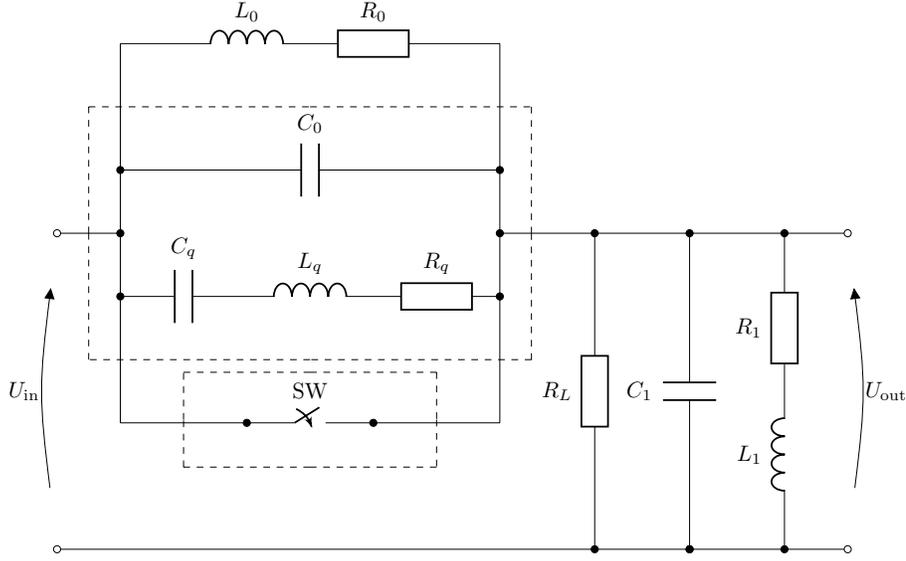}
\caption{Experimental circuit. The quartz is represented in a dashed box by a series RLC with $R_q$, $C_q$ and $L_q$ for the motional part and a parallel shunt capacitor $C_0$ (capacitance of the quartz plates). The analog switch is used to vary the load coupling resistor $R_L$. The switch and the printed circuit board have a parasitic capacitance $C_1$. We intentionally add two branches $L_0$,$R_0$ and $L_1$,$R_1$ to compensate $C_0$ and $C_1$ respectively.}
\label{fig:circuit_quartz}
\end{figure}

Since the storage time is limited by the intrinsic Q-factor of the resonator, we use a quartz as a widely available high quality component. In practice, a quartz oscillator is not simply a series RLC circuit. A simplified equivalent model can be represented by the 4-component circuit in fig.\ref{fig:circuit_quartz}. 
The mechanical vibration of the crystal is reproduced by a series RLC with $R_q$, $C_q$ and $L_q$, in parallel with a capacitance, $C_0$, which represents the electrical plates contact of the transducer \cite{bottom1982introduction}.

Because of the parallel shunt capacitance $C_0$, the quartz oscillator is far from an ideal band-pass filter as represented in fig. \ref{fig:circuit_equivalent}. As we will verify later, the shunt capacitance can be compensated by placing in parallel an inductor $L_0$ (including its winding resistance $R_0$) on the input side (Coilcraft 1812CS-472XJLC). Our commercial analog switch (SW in fig.\ref{fig:circuit_quartz}, Texas Instruments TS12A12511) have a parasitic capacitance $C_1$ (including the printed circuit board additional capacitance). On the output side, $C_1$ should be also compensated by placing an inductor $L_1$ (with winding resistance $R_1$ for the Coilcraft 1812CS-183XJLC) in parallel.

The compensation branches work as follows. The working frequency $\omega_r$ is the resonant frequency of the quartz (AEL Crystal X14M000000L001), typically $\omega_r=2\pi\times$14MHz in our case. We adjust the compensations $L_0$,$R_0$ and $L_1$,$R_1$ such as $R_0$,$L_0$,$C_0$ and $R_1$,$L_1$,$C_1$ are resonant at $\omega_r$  as well. In practice to adjust the resonant condition $L_0 C_0 \omega_r^2$ ($L_1 C_1 \omega_r^2$ respectively), we keep $L_0$ ($L_1$ resp.) fixed to its nominal condition and change $C_0$ ($C_1$ resp.) by placing an extra varying capacitor in parallel (not represented in fig.\ref{fig:circuit_quartz}). As we will see later, this compensation scheme is critical to approach the ideal situation of fig.\ref{fig:circuit_equivalent}.

\subsubsection{Circuit characterization}\label{bode}

The motional components of the quartz are estimated from the commercial specifications, $R_q=25\,\Omega$, $L_q=6.47$ mH, and slightly modified for the capacitance $C_q=19.985$ fF to reproduce the exact measured resonant frequency $\omega_r=2\pi\times$13.9964MHz close to the nominal value of 14MHz for $C_q\simeq20$ fF.

To fully characterize our circuit and to anticipate the pulse storage by dynamically controlling the coupling though the load $R_L$, we vary step by step $R_L$ from $100\, \Omega$ to $10\, \mathrm{k\Omega}$ in order to fit the different values of  $C_0$ and $R_0$   ($C_1$ and $R_1$ respectively) keeping $L_0=18\,\mu$H and $L_1=4.7\,\mu$H to their specification values. $C_0$ and $R_0$ are not precisely known ($C_1$ and $R_1$ respectively) because on the one hand we modify the capacitance to reach the resonant condition of the compensation branches and on the other hand the resistance $R_0$ ($R_1$ respectively) is the winding series resistance which depends on the frequency. That's why we use a fitting procedure to retrieve the different values.

\begin{figure}[htbp]
\centering
\includegraphics[width=.8\linewidth]{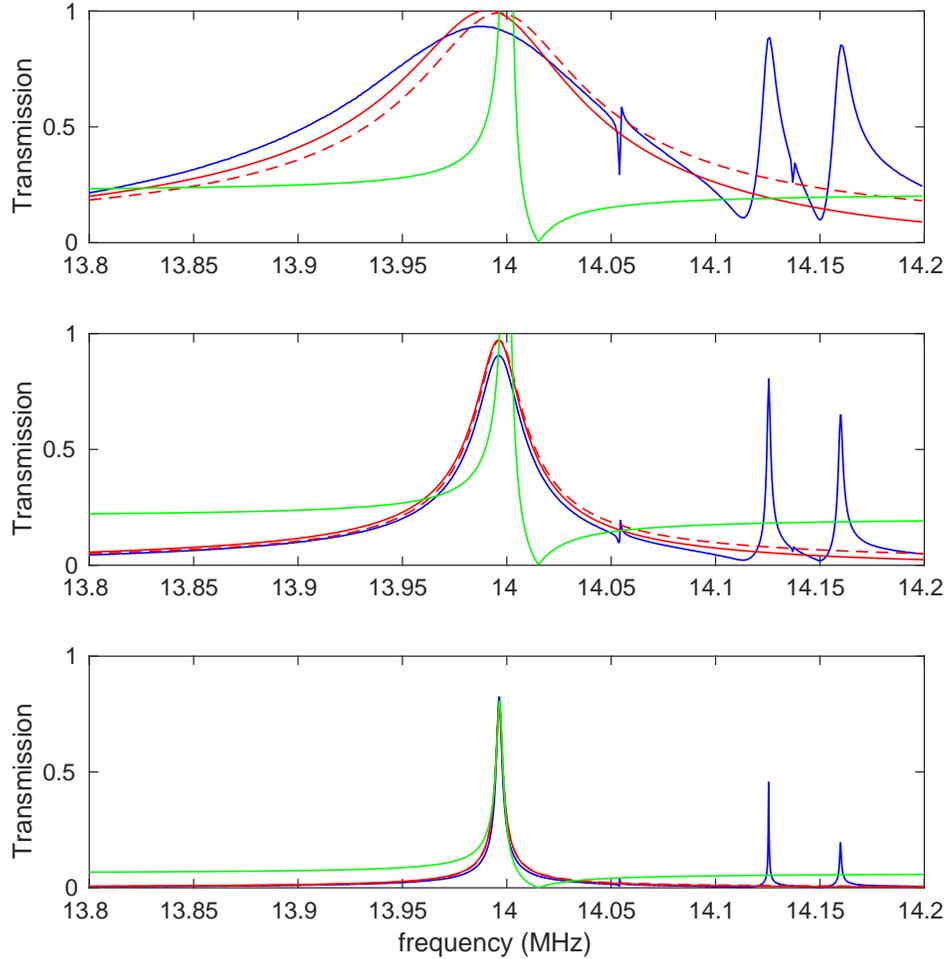}
\caption{Bode diagram (transmission) for $R_L =9.98\, \mathrm{k\Omega}$ (top),  $R_L = 1.02\, \mathrm{k\Omega}$ (middle) and $R_L = 100\, \mathrm{\Omega}$ (bottom). The blue curves are the experimental data. The red curves are the fits to data using the procedure described in the text with the circuit of fig.\ref{fig:circuit_quartz}. The red dashed curves are the fits to data with an simplified equivalent model described in \ref{section:equivalent_circuit} and represented in fig.\ref{fig:circuit_quartz_compensated}. This later doesn't account for the two parasitic resonances observed close to 14.15MHz. The green curves represent the expected spectra without the compensation branches ($L_0$,$R_0$ and $L_1$,$R_1$ removed).}
\label{fig:fit_compensations_minimize_figure}
\end{figure}

We vary $R_L$ from $100\, \Omega$ to $10\, \mathrm{k\Omega}$ step by step ($100\, \Omega$,  $200\, \Omega$,  $300\, \Omega$,  $540\, \Omega$,  $1.02\,  \mathrm{k\Omega}$,  $2\,  \mathrm{k\Omega}$,  $3\,  \mathrm{k\Omega}$,  $5.4\,  \mathrm{k\Omega}$ and  $9.98\,  \mathrm{k\Omega}$). We fit this ensemble of 9 spectra with $C_0$, $R_0$, $C_1$ and $R_1$ as free parameters. The result is plotted in red in fig.\ref{fig:fit_compensations_minimize_figure} where we have represented only three significant values of $R_L = 100\, \mathrm{\Omega}$, $1.02\, \mathrm{k\Omega}$  and $9.98\, \mathrm{k\Omega}$.

The agreement is imperfect especially at large $R_L$ values like $9.98\, \mathrm{k\Omega}$. The discrepancy is attributed to the additional resonances at 14.13 MHz and 14.16 MHz clearly appearing in the spectrum. They are not included in our model with a singly resonant $R_q$, $L_q$, $C_q$ oscillator. They overlap the main peak at 14 MHz producing an asymmetry in the central bandwidth at large $R_L$ (large bandwidth). From the fit, we obtain the values of $C_0=7.4$ pF, $R_0=53\, \mathrm{\Omega}$, $C_1=27$ pF and $R_1=37\, \mathrm{\Omega}$ consistent with the specifications of the components.

In conclusion, we obtain a band-pass filter centered at 14 MHz whose width can be controlled by varying the load $R_L$. Our compensated circuit behaves as a band-pass filter with a variable coupling. Our double compensation scheme with $L_0$,$R_0$ and $L_1$,$R_1$ is critical. As we see in fig. \ref{fig:fit_compensations_minimize_figure} (green curve), without the compensation branches, the real circuit would not behave as a band-pass filter. Before exploiting our dynamical filter for pulse storage, we propose to simplify further our analysis and show that the compensation branches can be easily modeled by a single equivalent resistor.

\subsubsection{Equivalent circuit}\label{section:equivalent_circuit}

Our analysis can indeed be simplified by noting that the compensations $L_0$,$R_0$  ($L_1$,$R_1$ resp.) in parallel with $C_0$ ($C_1$ resp.) form parallel RLC resonators. The complex impedance close to resonance is real and reads as $R_i^\mathrm{eq}=R_i Q_i^2$ (for $i=0,1$) where $\displaystyle Q_i=\frac{1}{R_i} \sqrt{\frac{L_i}{R_i}}$ is the Q-factor of the parallel R$_i$L$_i$C$_i$. For the fitted values $R_i$,$L_i$ and $C_i$, we obtain $R_0^\mathrm{eq}=45\, \mathrm{k\Omega}$ and  $R_1^\mathrm{eq}=4.8\, \mathrm{k\Omega}$.

\begin{figure}[htbp]
\centering
\includegraphics[width=.6\linewidth]{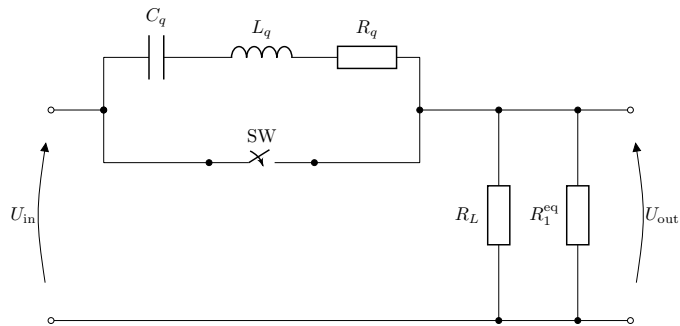}
\caption{Equivalent circuit with adapted compensation. As compared to fig. \ref{fig:circuit_quartz}, 
when properly adjusted $L_0$,$R_0$ compensate the parallel shunt capacitor $C_0$ (see text for details). The total impedance of the $L_0$,$R_0$, $C_0$ becomes negligible with respect to the quartz. On the output side, the branch $L_1$,$R_1$ compensating $C_1$ is equivalent to a resistor $R_1^\mathrm{eq}$ whose value should be included in parallel with $R_L$ as a total circuit load.
}
\label{fig:circuit_quartz_compensated}
\end{figure}

$R_0^\mathrm{eq}$ is much larger than the impedance of the quartz and can then be neglected completely. In other words, $C_0$ can be fully compensated and removed from the analysis. On the contrary, $R_1^\mathrm{eq}$ is comparable to the load resistance that we will use for the storage. This is not negligible and must be included in the resistive load as an effective load at the output. The equivalent circuit (see fig.\ref{fig:circuit_quartz_compensated}) is very similar to the ideal case of a variable band-pass (fig.\ref{fig:circuit_quartz}) where the output load $R_\mathrm{out}$ written $R_\mathrm{out}=R_1^\mathrm{eq}\parallel R_L$ is the parallel resistance of $R_1^\mathrm{eq}$ and $R_L$.

To characterize more accurately the effective load $R_1^\mathrm{eq}$, we simply reuse the fitting procedure detailed in section \ref{bode} with the equivalent circuit in fig.\ref{fig:circuit_quartz_compensated} and leave now $R_1^\mathrm{eq}$ as a single free fitting parameter. We obtain $R_1^\mathrm{eq}=4.3\, \mathrm{k\Omega}$ with the corresponding fits as red dashed lines in fig.\ref{fig:fit_compensations_minimize_figure} close to the expected $R_1 Q_1^2=4.8\, \mathrm{k\Omega}$. The agreement is also imperfect at large $R_L$ values but is quite satisfying for a single parameter fit. We conclude that our circuit is indeed equivalent to a band-pass filter as represented in fig.\ref{fig:circuit_quartz_compensated} despite the intrinsic complexity of the quartz. After our characterization in the spectral domain (Bode diagram), we now investigate the transient response of the circuit when the output load is varied dynamically to store an incoming pulse.

\section{Pulse storage though dynamical coupling}\label{sec:storage}

The ideal filter of fig.\ref{fig:circuit_quartz_compensated} with a variable output coupling can be used to store incoming RF photons. The term "variable coupling" should be precised in our case, first qualitatively. When the switch (SW in fig.\ref{fig:circuit_quartz_compensated}) is off, the quartz oscillator is coupled to the input and the output through the load $R_L$. It can be directly excited by the input pulse for example. By activating the switch (SW on), the quartz is looped back into itself. The electric excitation oscillates between the motional components $L_q$ and $C_q$ and is damped by $R_q$. This oscillation corresponds to the free running vibration of the quartz crystal. When the switch is released back off, the free running excitation is transferred to the output load. This is actually a general storage scheme that we translate and interpret with a RLC resonator  \cite{yanik2005stopping}. More quantitatively, the incoming pulse duration should be comparable to the inverse of the filter bandwidth with the output load $R_\mathrm{out}$ (switch off). The incoming bandwidth is then $\displaystyle \Gamma = \frac{R_\mathrm{out}}{L_q}$ ($R_q$ is negligible with respect to $R_\mathrm{out}$).  During the free oscillation (switch on), the storage time is limited by the Q-factor or the intrinsic damping time of the quartz which is $\displaystyle \frac{L_q}{R_q}$. In other words, the storage time-to-bandwidth product is given by $\displaystyle \frac{R_\mathrm{out}}{R_q}$. This latter is ultimately limited by the quartz Q-factor. This analysis is sufficient to derive the different orders of magnitude but it can be pushed one step further. How fast should be the switch off/on time ? It should be faster than the inverse of the bandwidth $\displaystyle \Gamma$. This simple statement should be kept in mind because the rapidity to dynamically control the filter is potentially a limiting factor of the storage bandwidth. The switch control should be designed accordingly.

We have shown that our circuit can be modeled by band-pass filter with a variable load. This latter can be formally described by a Lorentzian response for the transfer function (eq.\ref{Bode_second}) as expected for an ideal band-pass filter. In the case of an high-Q oscillator close to resonance $\omega \sim \omega_q$,  the response eq.(\ref{Bode_second}) can be simplified to the first order by 

\begin{equation}
\frac{\TFU{out}{\omega}}{\TFU{in}{\omega}} \simeq \frac{1}{1+2j\left(\omega-\omega_q\right)/\Gamma}
\end{equation}

The description in the spectral domain is meaningless if the load is varied or in other words if $\Gamma$ changes in time. So far we focus on the initial situation when $\displaystyle \Gamma = \frac{R_\mathrm{out}}{L_q}$. This allows to properly describe the transient response of the initial filter and guide our intuition at the storage step when $\Gamma$ is reduced to ideally zero and practically to its minimum value $\displaystyle \frac{R_\mathrm{q}}{L_q}$. We introduce the slowly varying complex envelopes $V_k(t)$  for $k=\left\{ \mathrm{in},\mathrm{out} \right\}$ as opposed to the rapidly oscillating waveforms $U_k(t)$ by

\begin{equation}
U_k(t)=V_k(t) \exp\left(j \omega t\right)
\end{equation}

So the transfer function compactly reads as

\begin{equation}
\frac{\TFV{out}{\omega}}{\TFV{in}{\omega}} = \frac{1}{1+2j\omega/\Gamma} \label{Bode_lorentzian}
\end{equation}

and fully describes the response of the filter when the switch is off. The output response depends on the temporal shape of the incoming pulse. This obviousness  actually raises the question of the storage efficiency. What is the optimized incoming shaped to maximize the storage efficiency ? The incoming pulse duration is qualitatively given by the bandwidth $\displaystyle \Gamma$. This doesn't tell anything about the exact pulse shape as we will discuss now.

\subsection{Optimal pulse shape and expected storage efficiency}\label{optimal}

\begin{figure}[htbp]
\centering
\includegraphics[width=.6\linewidth]{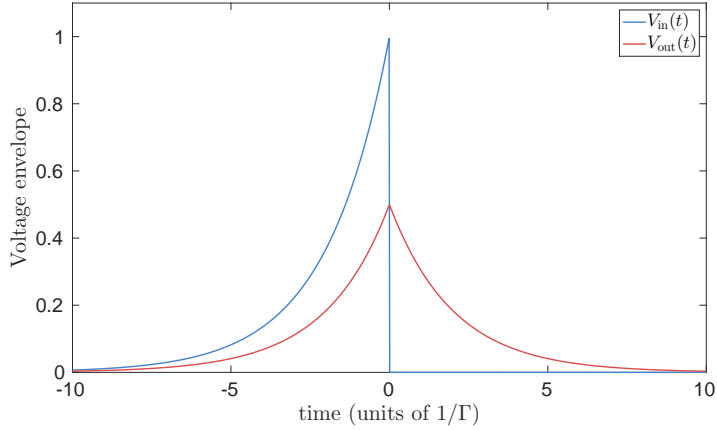}
\caption{Output pulse $V_\mathrm{out}(t)$ when the input $V_\mathrm{in}(t)$ is an exponentially rising pulse matching the Lorentzian resonator lifetime $1/\Gamma$.}
\label{fig:exp_in_out}
\end{figure}

\begin{figure}[htbp]
\centering
\includegraphics[width=.65\linewidth]{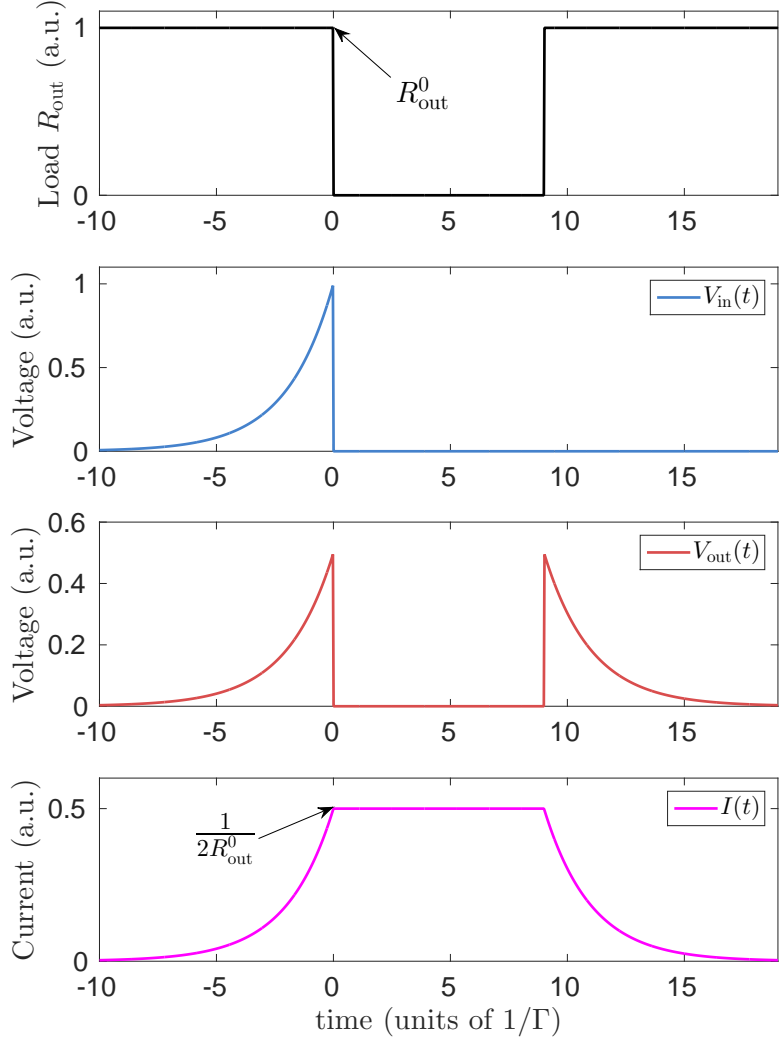}
\caption{Storage sequence of an exponentially rising pulse $V_\mathrm{in}(t)$ when the load is varied from its initial value $R_\mathrm{out}^0$ to zero (top). The evolution in fig. \ref{fig:exp_in_out} is interrupted and the output pulse $V_\mathrm{out}(t)$ is retrieved when switch is deactivated. We have also represented the current in the circuit.}
\label{fig:exp_in_out_storage}
\end{figure}

The question of the optimal pulse shape in the context of storage has been consistently addressed by many authors from different communities \cite{Gorshkov_optimal, Gorshkov_A, Gorshkov_B, 0295-5075-86-1-14007_ref18_QM, PhysRevA.83.063842_ref19_QM, Heugel2010_ref38_Cl, PhysRevB.84.014510_ref22_Cl, PhysRevLett.112.210501_Martinis}. Without going into details, one should simply remember that the optimization is based on time reversal arguments. In other words, the optimum in reached when the retrieved signal is the time-reversed of the incoming pulse corresponding to complex conjugate parts of the spectrum. This time symmetry is revealed by figs.\ref{fig:exp_in_out} and \ref{fig:exp_in_out_storage} where rising and decaying exponentials illustrate the optimal storage conditions as we will discuss in this section. More generally, for a given arbitrary transfer function, the optimization procedure becomes an integral equation (the convolution by the impulse response) that can be solved recursively even in sophisticated situations \cite{Gorshkov_optimal}. This time symmetry between signal and retrieval has been successfully exploited in dense atomic ensembles using the EIT scheme, allowing a significant efficiency gain \cite{Novikova}. In our case, the Lorentzian response as given by eq.\ref{Bode_lorentzian} is an archetype in that sense because the time reversed signal can be extracted easily. This is an exponentially rising pulse matching the resonator lifetime $\displaystyle 1/\Gamma$. More precisely for

\begin{equation}
V_\mathrm{in}(t)=H(-t) \exp\left(\Gamma t/2\right) \label{exp_growth}
\end{equation}
where $H()$ is the Heaviside step function. In the Fourier space
\begin{equation}
\TFV{in}{\omega}=\frac{2}{\Gamma}  \frac{1}{1-2j\omega/\Gamma}
\end{equation}
and for the output (eq.\ref{Bode_lorentzian})
\begin{equation}
\TFV{out}{\omega}=\frac{2}{\Gamma}  \frac{1}{1+4\omega^2/\Gamma^2}
\end{equation}
Back to the time domain, the output pulse can be decomposed into a rising and a decaying exponential as

\begin{equation}
V_\mathrm{out}(t)=\frac{1}{2} \exp\left(-\Gamma |t|/2\right) = \frac{H(-t)}{2} \exp\left(\Gamma t/2\right) +  \frac{H(t)}{2} \exp\left(-\Gamma t/2\right)\label{exp_dec}
\end{equation}

We have represented $V_\mathrm{in}(t)$ and $V_\mathrm{out}(t)$ in fig.\ref{fig:exp_in_out}.

The decaying part $\displaystyle \frac{H(t)}{2} \exp\left(-\Gamma t/2\right)$ is indeed the time reserved of $V_\mathrm{in}(t)$. So the incoming pulse shape will maximize the efficiency. The fact that $V_\mathrm{out}(t)$ is not only composed of a time reversed copy of $V_\mathrm{in}(t)$ but also of the rising exponential $ {H(-t)} \exp\left(\Gamma t/2\right)$ may be a source of confusion. This latter won't be stored because it is part of the excitation pulse as opposed to the decaying exponential which is given by the free oscillation of the quartz after the excitation. This exponential decay is sometimes called a free induction decay in nuclear magnetic resonance \cite{PhysRev.77.297.2} or in the optical domain \cite{PhysRevA.6.2001}. It is then relevant to compare $V_\mathrm{in}(t)$ and $\displaystyle \frac{H(t)}{2} \exp\left(-\Gamma t/2\right)$ exclusively (neglecting the other part $\displaystyle \frac{H(-t)}{2} \exp\left(\Gamma t/2\right)$) to confirm that the input and the stored part are indeed time-reserved corresponding to the optimal storage. The prefactor $\displaystyle\frac{1}{2}$ gives the maximum retrieval efficiency $\displaystyle 25\%$ when energies (as the square of the envelopes) are compared. The maximum value of  $\displaystyle 25\%$ is given by the filter transfer function (eq. \ref{Bode_lorentzian}). This can reach $100\%$ for a asymmetric resonator as shown by Bader {\it et al.} with an optical cavity \cite{1367-2630-15-12-123008}. In that case, the input and output ports of the resonator are the same as opposed to our two-port band-pass filter (fig. \ref{fig:circuit_equivalent}). As a consequence, the input and output pulses are exactly time-reversed with the same amplitude \cite[fig.2]{1367-2630-15-12-123008}. We here focus on the band-pass filter configuration thus we are intrinsically limited to $\displaystyle 25\%$.

To evaluate the expected efficiency, our time reversal analysis is based on the static case when the output load $R_\mathrm{out}$ is fixed. This static analysis can be extended when a dynamical control of the load is applied at the storage step. When the switch is activated, the quartz is looped back into itself. This correspond to zeroing the output load. The quartz freely oscillates. When the switch is deactivated at the retrieval step, the energy is released into the load. In other words, the switch interrupts and resumes the evolution depicted in fig.\ref{fig:exp_in_out} (static case). The storage sequence is represented in fig.\ref{fig:exp_in_out_storage}.

Fig.\ref{fig:exp_in_out_storage} can be seen as an interrupted version of fig.\ref{fig:exp_in_out} when the switch is activated ($R_\mathrm{out}$ goes to zero). During the interruption $R_\mathrm{out}=0$, the output voltage $V_\mathrm{out}(t)$ is zero because the switch acts as a short-cut. The current going through the circuit is simply  $\displaystyle I(t)=\frac{V_\mathrm{out}(t)}{R_\mathrm{out}}$ when the switch is off. $ I(t)$  is a continuous function so the current keeps a finite value $\displaystyle \frac{1}{2R_\mathrm{out}^0}$ during the storage period. Our qualitative analysis completely neglects the intrinsic damping time of the quartz due to $R_q$. If the damping time is comparable to the storage time, $V_\mathrm{out}(t)$ and  $I(t)$ will decay accordingly during the free evolution period leading to a reduced retrieval amplitude. We will now implement this situation using the experimental circuit presented in \ref{section:experimental_realization}.

\subsection{Experimental efficiency and storage time}\label{experimental}
We start with a large load resistor to ensure the largest initial bandwidth for the filter. We take $R_L = 9.98\, \mathrm{k\Omega}$ corresponding to fig.\ref{fig:fit_compensations_minimize_figure} (top). The output load is initially $R_\mathrm{out}^0=R_1^\mathrm{eq}\parallel R_L$ with $R_1^\mathrm{eq}=4.3\, \mathrm{k\Omega}$ derived from the fitting procedure described in \ref{bode}. This latter is actually limited by the impedance $ R_1^\mathrm{eq}$ of the output compensation branch. So the initial bandwidth is $\displaystyle \Gamma = \frac{R_\mathrm{out}}{L_q}=2\pi\times$74.2kHz with the fitted value of $L_q=6.47$ mH. The optimal shape is an exponentially rising pulse given by the eq.(\ref{exp_growth}) with duration $\displaystyle \frac{2}{\Gamma}=4.29\mu s$ that we program with an arbitrary waveform generator (AWG) WavePond DAx22000-8M (Chase Scientific).

\begin{figure}[htbp]
\centering
\includegraphics[width=.8\linewidth]{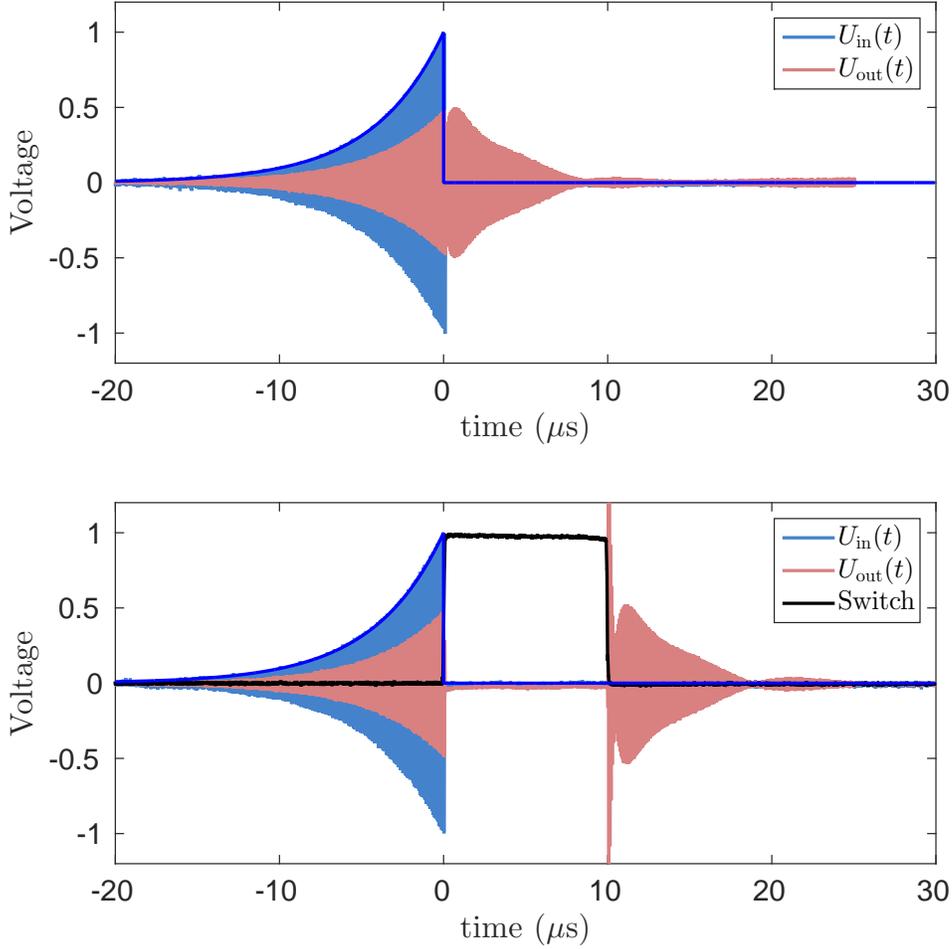}
\caption{Top: Output pulse $U_\mathrm{out}(t)$ (in red) when the input $U_\mathrm{in}(t)$ (in blue) is an exponentially rising pulse of duration $4.29\mu s$. The carrier frequency is $\omega_r=2\pi\times$13.9964MHz. We have underlined the incoming pulse envelope $V_\mathrm{in}(t)$ in dark blue. Bottom: When the switch (in black) is activated at time $0$, the pulse is stored. The retrieval is triggered at $t=10\mu s$ when the switch is deactivated following the scheme of fig. \ref{fig:exp_in_out_storage}.}
\label{fig:SL_St_exp}
\end{figure}

Following our description in \ref{optimal}, we first characterize the temporal response of the circuit (without activating the switch) and compare it with the expected output envelope in fig.\ref{fig:exp_in_out}. For a carrier frequency of $\omega_r=2\pi\times$13.9964MHz, we have represented the output response in fig.\ref{fig:SL_St_exp}(top).

The output pulse $U_\mathrm{out}(t)$ is slightly distorted from the purely exponential components of eq.\ref{exp_dec}. This is irredeemably associated to the deviation from the Lorentzian shape of the Bode diagram in fig.\ref{fig:fit_compensations_minimize_figure}(top), and more precisely to the global asymmetry of the shape including the side resonances at 14.13 MHz and 14.16 MHz. Despite a slight qualitative deviation, the output amplitude is indeed $\frac{1}{2}$ of the input as expected from eq.\ref{exp_dec}.

We now turn to the storage sequence described in fig.\ref{fig:exp_in_out_storage} by activating the switch thus effectively varying the load impedance from its initial value $R_\mathrm{out}^0$ to zero. The result is represented in fig. \ref{fig:SL_St_exp}(bottom). The switch is activated during $\tau=10\mu s$ corresponding to the storage duration. Despite a transient response at $t=10\mu s$, the delayed part of $U_\mathrm{out}(t)$ is indeed stored and released without major distortions. This visual impression validates our interpretation developed in \ref{optimal} and directly derived from the static case (no switch). The evolution in the static case is interrupted by the switch leading to an effective storage sequence.

The output voltage is again close to $\frac{1}{2}$ of the input as expected from eq.\ref{exp_dec}. The efficiency should be close to 25\% (as the square of the envelopes). The storage efficiency $\eta$ can be evaluated more precisely by extracting the pulse envelope $V_\mathrm{out}(t)$ from the waveform $U_\mathrm{out}(t)$ and by calculating the integral of the envelope squared after the retrieval time $\tau$ ($\tau=10\mu s$ in that case) \footnote{The envelope is numerically given by the amplitude of the analytic signal (Hilbert transform).}:

\begin{equation}
\eta=\int_\tau^\infty V_\mathrm{out}(t)^2 \mathrm{d}t \label{eta}
\end{equation}

We obtain 26.6\% for a storage $\tau=10\mu s$. This is larger than the maximum expected of 25\%. This is likely due to the transient response of the switch at the retrieval time (sharp peak visible at $10\mu s$ in fig. \ref{fig:SL_St_exp},bottom). This deserves further investigation but is beyond the scope of our demonstration.

\begin{figure}[htbp]
\centering
\includegraphics[width=.8\linewidth]{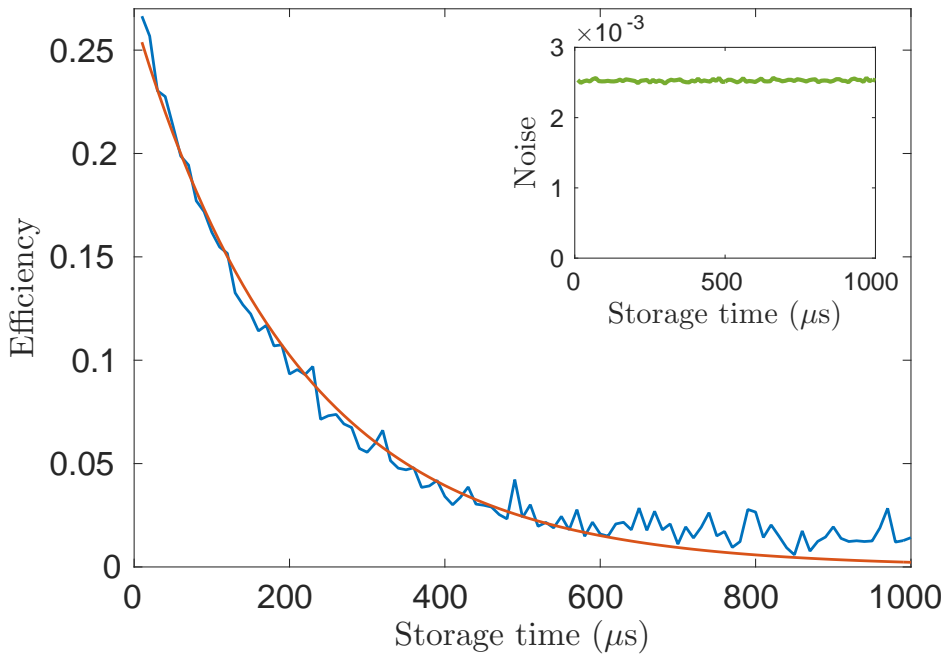}
\caption{Storage efficiency as a function of the storage time. The efficiency decays exponentially with the storage time whose decay constant is fitted to $209\mu s$. This is our memory lifetime. Inset: Noise floor evaluated operating the circuit without the input signal.}
\label{fig:Trait_EfficaciteVsTretention_octave_figure}
\end{figure}

To further characterize our memory, we make vary the storage time $\tau$ from $10\mu$s to $1$ms (fig. \ref{fig:Trait_EfficaciteVsTretention_octave_figure}).

The efficiency decays exponentially with a constant of $\displaystyle \frac{1}{\Gamma_\mathrm{exp}}=209\mu s$ (fig. \ref{fig:Trait_EfficaciteVsTretention_octave_figure}). This is our memory lifetime. The storage duration is actually limited by the intrinsic losses of the quartz. The lifetime can be estimated from the fitted values in \ref{bode} and is given by $\displaystyle \frac{1}{\Gamma_q}=\frac{L_q}{R_q}=259\mu s$. The slight difference with the measured $\displaystyle \frac{1}{\Gamma_\mathrm{exp}}=209\mu s$ is well explained by the internal switch resistance ($\sim 5 \Omega$ nominally).

The input envelope $V_\mathrm{in}(t)$ duration is $\displaystyle \frac{2}{\Gamma}=4.29\mu s$ or alternatively $\displaystyle \frac{1}{\Gamma}=2.15\mu s$ if the energy is considered (as $V_\mathrm{in}(t)^2$). From that, we can define the time-to-delay product of the memory as $\displaystyle \frac{\Gamma_\mathrm{exp}}{\Gamma}=98$. This latter is theoretically limited by the Q-factor of the quartz. But in practice, it is limited by the impedance $ R_1^\mathrm{eq}$ of the output compensation branch. A larger value of $ R_1^\mathrm{eq}$ is desirable. This is a clear limitation of the quartz whose parallel shunt capacitance needs to be compensated in our scheme.

We also evaluate the noise level, as it would be done in a storage experiment, by operating the circuit without the input signal (fig. \ref{fig:Trait_EfficaciteVsTretention_octave_figure}, inset). The background level is due to the transient response of the analog switch that we use to dynamically control the coupling constant. As a consequence, the noise floor doesn't depend on the storage time and is essentially constant $\sim$ 0.25\%.

\subsection{Storage of arbitrary input shapes}

\begin{figure}[htbp]
\centering
\includegraphics[width=.8\linewidth]{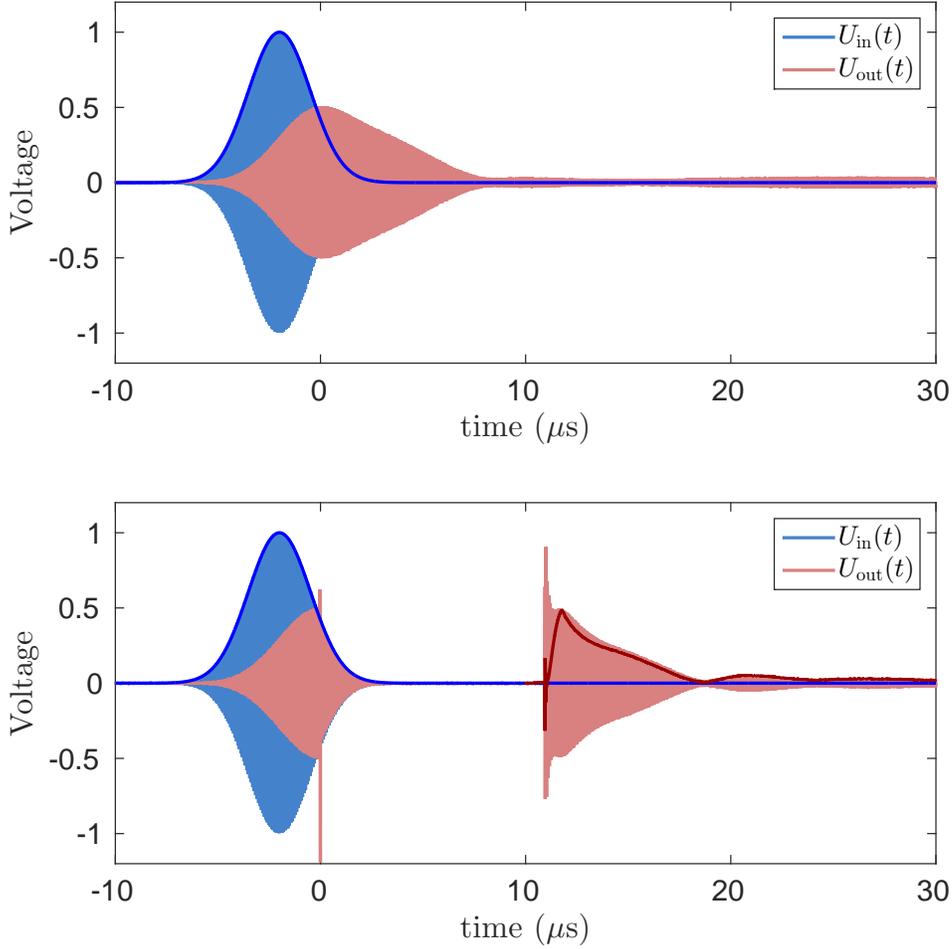}
\caption{Top: Output pulse $U_\mathrm{out}(t)$ (in red) when the input $U_\mathrm{in}(t)$ (in blue) is an exponentially rising pulse of duration $\sigma=1.52\mu s$. We have underlined the incoming pulse envelope $V_\mathrm{in}(t)$ in dark blue. Bottom: When the switch (in black) is activated at time $0$ after a delay $\sigma$, the pulse is stored. The retrieval is triggered at $t\simeq11\mu s$ when the switch is deactivated. The outgoing pulse envelope $V_\mathrm{out}(t)$ is underlined in dark red.}
\label{fig:SL_St_gauss}
\end{figure}

In order to illustrate the possibility to store different shapes than the exponentially rising pulse (fig. \ref{fig:SL_St_exp}), we apply our scheme to a Gaussian pulse in fig.\ref{fig:SL_St_gauss}.

We choose a Gaussian as
\begin{equation}
V_\mathrm{in}(t)= \exp\left(\frac{ -t^2}{2\sigma^2}\right) \label{gauss}
\end{equation}
whose duration $\sigma=1.52\mu s$ has the same $\displaystyle \frac{1}{e}$ duration ($4.29\mu s$) of the exponentially rising pulse of fig. \ref{fig:SL_St_exp}. The output pulse (fig.\ref{fig:SL_St_gauss}, top) is significantly distorted. When the switch is activated (fig.\ref{fig:SL_St_gauss}, bottom), a small fraction of the incoming Gaussian is actually stored. The input signal and the retrieval have very different shapes. They are not simply related by a time symmetry as this is the case for rising and decaying exponentials. As a symptom of this time reversal breaking, the efficiency is lower: approx. 15\% as the integral of the intensity envelop (square of the red line envelope in fig. \ref{fig:SL_St_gauss}, bottom).

The trade-off that should be found between the incoming pulse duration and the slow-light pulse observed in fig.\ref{fig:SL_St_gauss} (top) is a recurrent concern of the experimentalists doing slow and stopped-light. The incoming pulse and as a consequence the retrieval are typically clipped because of the switch activation. This clipping as illustrated by fig. \ref{fig:SL_St_gauss} (bottom) can be alternatively interpreted as an incomplete fitting of the spatially extended pulse whose raising and falling tails leak out of the storage medium. This realization for Gaussian pulses aims at reinforcing the analogy with EIT storage as we will finally discuss in section \ref{stopped_light}.

To explore a bit further the feature of our memory, we will now investigate its coherent nature.

\section{Coherent memory}\label{coherent}

The coherent character of a memory is crucial for quantum storage. Our rudimentary storage scheme actually preserves the coherence between input and output. This is not guaranteed because the storage sequence by definition includes a free evolution period. Does this quartz free oscillation preserves the phase ? This can be verified by simply varying the input phase and show that the output follows. This a clear advantage of the RF domain where the field oscillations can be easily recorded as compared to optics (intensity measurement). We additionally make interfere two independent memories (as we would do in the optical domain) and show that well contrasted interference fringes can be obtained. 

\subsection{Input/output phase relation}\label{in_out_phase}

\begin{figure}[htbp]
\centering
\includegraphics[width=.5\linewidth]{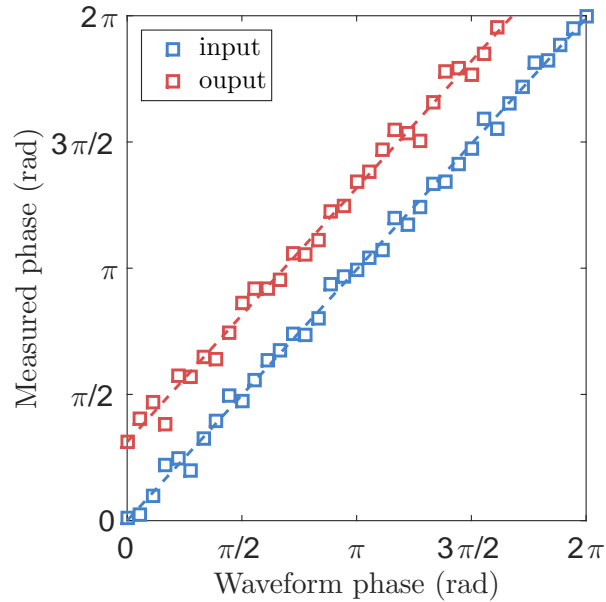}
\caption{Input/ouput phase relation. The measured phase of the input allows to verify that the actually programmed phase is indeed reproduced by the AWG. Dashed line are linear regressions. The measured output phase follows accordingly with an offset of 0.32$\pi$ (see text).}
\label{fig:trait_acq_var_phase_resume}
\end{figure}

With the storage sequence presented in fig.\ref{fig:SL_St_exp}(bottom), we change the phase of the input pulse $U_\mathrm{in}(t)$ through the AWG. The phase of the retrieved signal should follow accordingly as plotted in fig.\ref{fig:trait_acq_var_phase_resume}.

After programming the waveform with a given phase, we first verify that it  is indeed reproduced by the AWG. As in fig.\ref{fig:SL_St_exp}(bottom), we record  the input pulse $U_\mathrm{in}(t)$ from the oscilloscope, calculate its Fourier transform and extract the phase at $\omega_r=2\pi\times$13.9964MHz (measured input phase in blue in fig.\ref{fig:trait_acq_var_phase_resume}). We apply the same procedure with the output  $U_\mathrm{out}(t)$, but we only consider the retrieval by windowing the pulse after the storage time $\tau$. This gives the measured output phase (red in fig.\ref{fig:trait_acq_var_phase_resume}). The output phase follows the input as expected for a coherent storage.

There is a constant offset between the input and output phase of 0.32$\pi$. This is not expected. This could be explained by an inaccurate evaluation of the resonant frequency $\omega_r$. In that case, because of the detuning between the excitation frequency and the free running oscillation of the quartz, a phase mismatch is accumulated during the storage step. A 8kHz detuning would produce an offset of 0.32$\pi$ for the storage time used in this experiment ($\tau = 20.54\mu$s). Exploring the origin of this mismatch deserves further investigation. Whatever it is, the phase offset is well defined and constant so the storage sequence is indeed coherent. To investigate the reliability and the reproducibility of our scheme, we also make interfere two independent memories.

\subsection{Interference between memories}

\begin{figure}[htbp]
\centering
\includegraphics[width=.6\linewidth]{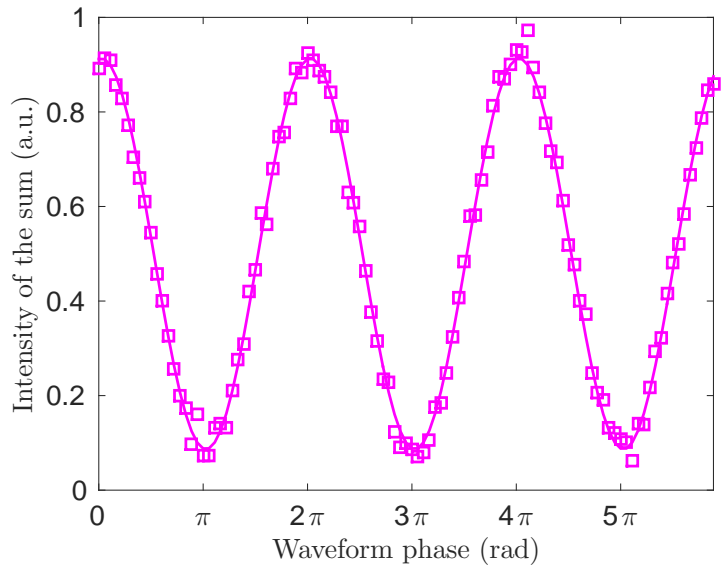}
\caption{Interference pattern of two memories when the input waveform phase of the second is varied. The storage time is $\tau = 20.54\mu$s in that case.}
\label{fig:MesureInterferencesVsPhase_resume}
\end{figure}

To explore further the coherent nature of the retrieval, we duplicate the experiment using the same components and apply the same compensation procedure. For the second memory,  the input waveform is phase shifted by a constant value that we vary from 0 to $6\pi$ (called waveform phase in fig.\ref{fig:MesureInterferencesVsPhase_resume}). The two retrieved signals should be phase shifted accordingly. To verify this assumption, we record both output waveforms by windowing the retrieved pulse after the storage time $\tau$. The sum of the field is done {\it in silico} as a post-processing stage but could be done analogically with standard RF components. The interference pattern in fig. \ref{fig:MesureInterferencesVsPhase_resume} is obtained after integrating the square of fields sum (the intensity).
 
We observe well contrasted interference fringes (91\% contrast) between the memories. With respect to the results of fig.\ref{fig:trait_acq_var_phase_resume}, the interference pattern doesn't give more information about the coherent character of one memory. This actually tells us that our scheme can be scaled from one circuit to a second one with a 91\% fidelity without special care on the reproducibility of the procedure. This figure of merit can certainly be improved even if this is not the main focus of the present demonstration. The interference pattern is nonetheless a strong and familiar evidence of coherence.

\section{Stopped-light in a quartz: alternative interpretation}\label{stopped_light}

Our modest setup cannot be compared to the advanced demonstrations of quantum storage. We share nevertheless a common motivation (coherent storage) and the same fundamental background. In the introduction, we have placed our storage scheme in a more general framework. Storage means controlling dynamically the coupling constant of an oscillation whatever is its physical nature. This truism is particularly intuitive to describes storage of light (resp. RF) in a cavity by controlling the reflectivity of one input mirror (resp. input RF coupler). The input mirror (resp. coupler) is described as a gate which can be open or closed to store and release the signal. This naive description is somehow limited because as we discussed the pulse duration and shape have to be optimized by taking into account the coupled oscillator lifetime (see section \ref{optimal}).

A resonant cavity considered as an oscillator is a special case in that sense. The pulse bandwidth has to be adjusted to the resonator lifetime, this latter being varied dynamically. A Fabry-Perot resonator, well known by opticians, is actually composed of many modes separated by the free-spectral range (FSR). In our description, we consider only one single mode, so by definition the bandwidth is smaller than the FSR. Or in other words, the round-trip time in the cavity is always the shortest timescale. The fact that a cavity is composed of many modes clearly offers a degree of freedom by introducing a second time scale in the problem. This is a potential source of richness for light storage but is beyond the scope of our approach focusing on a single resonant mode.

To broaden our vision, we briefly remind in conclusion that the stopped-light experiments based on EIT, extensively used for quantum storage of light, can be formally interpreted as a dynamical coupling to an effective oscillator. The analogy comes from the definition of the slow-light transfer function. This latter can be described by a complex propagation constant $\gamma_p(\omega) $ giving an input/output relation for the electric field that we write again as complex envelopes in the spectral domain as $\TFV{k}{\omega}$  ($k=\left\{ \mathrm{in},\mathrm{out} \right\}$). The propagation reads as

\begin{equation}
\frac{\TFV{out}{\omega}}{\TFV{in}{\omega}} = \exp \left(\gamma_p(\omega) L\right) \label{propag}
\end{equation}
$L$ is the length of the medium. The propagation constant $\gamma_p(\omega)$ is directly proportional to the atomic susceptibility of a three-level $\Lambda$-system \cite{fleischhauer2000dark, RevModPhys.77.633}. A narrow transparency window is open by applying the so-called control field (with a time varying intensity $\Omega(t)^2$). This latter is equivalent to our switch. When the transparency window (defining the input bandwidth) is sufficiently narrow as compared to the absorption profile (width $\Gamma_\mathrm{abs}$), the susceptibility can be written to the first order as  \cite{fleischhauer2000dark}

\begin{equation}
\gamma_p(\omega) = -\frac{\alpha}{2} \frac{2j\omega/\Gamma_\mathrm{EIT}}{1+2j\omega/\Gamma_\mathrm{EIT}} \label{propag_constant}
\end{equation}
$\alpha$ is the absorption coefficient. The width of the transparency window is given by the expression $\displaystyle \Gamma_\mathrm{EIT} = \frac{\Omega(t)^2}{2\Gamma_\mathrm{abs}}$.

A purely Lorentzian response is obtained only at low optical depth leading to

\begin{equation}
\frac{\TFV{out}{\omega}}{\TFV{in}{\omega}} \simeq \left(1-\frac{\alpha L}{2}\right)+\frac{\alpha L}{2}  \frac{1}{1+2j\omega/\Gamma_\mathrm{EIT}} 
\end{equation}
The Lorentzian profile appears as an archetype to describe the transfer function as in eq.(\ref{Bode_lorentzian}). For the EIT, the width $ \Gamma_\mathrm{EIT}$ is varied dynamically by an active control of the intensity $\Omega(t)^2$. $1/ \Gamma_\mathrm{EIT}$ scales the group delay of the so-called dark-state polariton \cite{fleischhauer2000dark}. In our case, the group delay is given by  $\displaystyle 1/\Gamma = \frac{L_q}{R_\mathrm{out}}$. This later goes to infinity when the output load goes to zero. Taking the terminology of stropped-light, the group velocity tends to zero, thus freezing the evolution of the polariton. The group velocity is not really a relevant figure of merit because the storage dynamics is fully described by the different time scales whatever is the exact physical length of the medium. The optical depth comes into play but it is still an extensive quantity. The length of the medium is to some extend arbitrary as soon as the total optical depth is well-defined as a dimensionless parameter.

At a large optical depth, the response is not a simple Lorenzian, but the exponential of a Lorentzian (eq. \ref{propag}). The physical description is qualitatively the same but with a different transfer function of variable width. The optimum pulse shape deviates from a rising exponential as a consequence. The optimization procedure which is trivial for a Lorentzian should be reconsider with the universal approach based on time reversal symmetry \cite{Gorshkov_optimal}.

There is no equivalent of the optical depth in our case because we consider a single oscillator. As a consequence, we can store only a single bit or a single temporal mode in the circuit. A very large optical depth allows to delay and finally store a train of pulses. This is a clear superiority of an atomic ensemble which can be seen as a stack of oscillators. The optical depth scales the multimode capacity of the memory as opposed to a single oscillator whose capacity in terms of qubit is one by definition. The extension of this work to an ensemble of coupled RLC oscillators is a perspective to explore further the analogy between RLC circuit and atoms \cite{scheuer2005coupled,cromieres2017slow}. As we already discussed, breaking the time-to-bandwidth product is crucial for storage. This is mostly due to a rapid variation of the coupling constant, or equivalently of the group delay, independently of the optical depth.

\section*{Conclusion}

We have implemented a coherent memory using a quartz as a high-Q resonator. The simplest storage scheme is achieved by varying the load, which is acting as an effective coupler. We obtain an efficiency of 26\% by implementing the shape optimization strategy that have been developed for light storage in atomic vapors. This is larger than the theoretical maximum of  25\% in our forward configuration. The slight discrepancy is explained by the transient response of the analog switch that we use to dynamically control the coupling constant. This would be associated to noise in a quantum storage scheme. The coherent character of our memory is verified by making interfere two storage devices. Without special care on the reproducibility, we obtain an interference fidelity of 91\%.

Our motivation is essentially fundamental. By applying different recipes from different communities (optics, atomic physics in the RF or optical domain) to a modest experimental setup, we would like to emphasize the essence of light storage in a material medium (independently of its physical reality). In its simplest version, a memory is based on a dynamical variation of the coupling constant to break down the time-to-bandwidth product.

\section*{Acknowledgments}

This work received funding from the national grant ANR DISCRYS (ANR-14-CE26-0037-02), from Investissements d'Avenir du LabEx PALM ExciMol and ATERSIIQ (ANR-10-LABX-0039-PALM). We thank E. Gozlan for his technical assistance.


\section*{References}
\bibliographystyle{iopart-num}
\bibliography{quartz_bib}

\end{document}